%% file: thebes.tex
\def\year{2021}\relax
\documentclass[letterpaper]{article} 
\usepackage{aaai21}  
\usepackage{times}  
\usepackage{helvet} 
\usepackage{courier}  
\usepackage[hyphens]{url}  
\usepackage{graphicx} 
\usepackage{natbib}  
\usepackage{caption} 
\usepackage{xcolor}

\usepackage{amsmath,amsfonts,amssymb}

\usepackage{subcaption}
\title{
How to Not Get Caught When You Launder Money on Blockchain?
}

\author {
    Cuneyt G. Akcora\textsuperscript{\rm 1}, Sudhanva Purusotham\textsuperscript{\rm 2}, Yulia R. Gel\textsuperscript{\rm 2}\\
    Mitchell Krawiec-Thayer\textsuperscript{\rm 3}, Murat Kantarcioglu\textsuperscript{\rm 2}  \\
}
\affiliations {
    \textsuperscript{\rm 1}University of Manitoba, 
   \textsuperscript{\rm 2} UT Dallas,  \textsuperscript{\rm 3} Monero Research Lab\\ 

    65 Chancellor Circle\\
    Winnipeg, MB, Canada R3T 2N2\\
    cuneyt.akcora@umanitoba.ca, \{Sudhanva.Purushotham, ygl, muratk\}@utdallas.edu, mitchell@insightfellows.com 

}
 
\begin{document}

\maketitle

\begin{abstract}
The number of blockchain users has tremendously grown in recent years. As an unintended consequence, e-crime transactions on blockchains has been on the rise. Consequently, public blockchains have become a hotbed of research for developing AI tools to detect and trace users and transactions that are related to e-crime.

We argue that following a few select strategies can make 
money laundering on blockchain virtually undetectable with most of the currently existing tools and algorithms. 
As a result, the effective combating of e-crime activities involving cryptocurrencies requires the development of novel analytic methodology in AI.
\end{abstract}

\section{Introduction}

Cryptocurrencies have emerged as an important rapidly evolving financial tool that allows for cross-border transactions without requiring a central trusted party. Unfortunately, the pseudo-anonymity of cryptocurrencies also continues to facilitate various forms of fraudulent activities.  Sellers and buyers of illicit goods are matched with ease, and payments are quickly executed. However,  contrary to fiat currencies,  where law enforcement agencies have access to a wide arsenal of analytic methods for investigation of various criminal activities, from money laundering to drug dealership to human trafficking, methods for detecting crime on the blockchain are yet in their infancy~\cite{foley2019sex, zhou2020market}. As a result, large amounts of illegal transactions with cryptocurrencies remain unidentified.  Given the complexity of the transaction patterns used by illicit networks,  simple heuristic-based techniques do not allow for reliable detection of suspicious blockchain activities and efficient implementation of Anti-Money Laundering (AML) rules. 
 
 To address this challenge, there have been recently proposed various AI techniques, e.g., \cite{deng2019fraudjudger, akcora2020bitcoinheist,lee2019cybercriminal}. 
 In addition, in September 2020 the U.S. Government solicited proposals for AI solutions in detecting money laundering activities on cryptocurrencies~\cite{irs}. 
 
 In this paper, we identify multiple techniques that can be used by the money launderers to hide illicit cryptocurrency transactions and circumvent existing AI-based money laundering techniques. Our insights, however, are intended not to assist criminals to obfuscate their malicious activities but to motivate the AI community for further research in this
area of critical societal importance and, as a result, to advance the state of art methodology in the AI-based cryptocurrency money laundering detection. 
 
 In the remainder of this paper, we first identify techniques that can be used by criminals to launder money on blockchain and then empirically show why these techniques would be effective in hiding patterns used by potential AI models.

\section{Related Work} 
 
The success of Bitcoin~\cite{nakamoto2008bitcoin} has encouraged hundreds of similar digital coins~\cite{tschorsch2016bitcoin}. The underlying Blockchain technology has been adopted in many applications.  With this rapidly increasing activity, there have been numerous studies analyzing the blockchain technology from different perspectives.   

The earliest results aimed at tracking the transaction network to locate coins used in illegal activities, such as money laundering and blackmailing \cite{androulaki2013evaluating,ober2013structure}. These findings are known as the taint analysis~\cite{di2015bitconeview}.  

Bitcoin provides pseudo-anonymity: although all transactions are public by nature, user identification is not required to join the network. Mixing schemes~\cite{maxwell2013coinjoin,ruffing2014coinshuffle} exist to hide the flow of coins in the network. However, as shown in~\cite{meiklejohn2013fistful}, 
 some Bitcoin payments can be traced. As a result, obfuscation efforts~\cite{narayanan2017obfuscation} by malicious users have become increasingly sophisticated. 

In ransomware analysis, Montreal~\cite{paquet2018ransomware}, Princeton~\cite{huang2018tracking} and Padua~\cite{conti2018economic} studies have analyzed networks of cryptocurrency ransomware and found that hacker behavior can aid identification of undisclosed ransomware payments. Datasets of these three studies are publicly available. 

Early studies in ransomware detection use decision rules on amounts and times of known ransomware transactions to locate undisclosed ransomware (CryptoLocker) payments~\cite{liao2016behind}. More recent studies are joint efforts between researchers and blockchain analytics companies~\cite{huang2018tracking,weber2019anti}. However, these studies do not disclose features used in their AI models.

Darknet market vendor identification has been considered by~\citet{tai2019adversarial} where features, such as shared email addresses, are extracted from market profiles and used in address classifiers.  \citet{lee2019cybercriminal} perform a taint analysis for addresses and track cryptocurrency payments to online exchanges. We glean insights from both articles and create a set of rules that militate against such analysis.

\section{Sources of Dark Coins}

\subsection{Darknet Markets}
Darknet markets are online sites that can only be accessed by privacy-enhancing tools such as TOR and I2P~\cite{market3}. Markets allow vendors to create online stores and match buyers to the vendors. The merchandise range from real passports to heroin and guns. The market employs an escrow service where a buyer is given a one-time-use address to pay for a good. After the payment, the market instructs the vendor to mail the goods physically. Upon the delivery (from abroad as well), the market releases coins to the vendor's account, which can cash out the coins. The escrow process can take around 6 days, and researchers found that in the early days (i.e., 2014) the release of coins could affect Bitcoin price~\cite{janze2017cryptocurrencies}. 
The market needs to foster trust between vendors and buyers to convince buyers to part with their coins. However, occasionally, the market defrauds vendors and disappears with coins, or similarly, vendors defraud buyers. 

Cryptocurrencies are ideal venues for Darknet markets because payments can be made anonymously and received anywhere in the world. Starting with the Silk Road in 2011, Darknet markets have been gathering ever-growing amounts where illicit goods, such as fake passports and guns, are transacted for cryptocurrencies. Law enforcement has seized multiple Darknet markets, but the ease of setting up a new market has turned this into a cat-and-mouse game.  Typically, Bitcoin and Monero are used for Darknet payments. In multiple cases, the seized markets' data have been made public~\cite{paquet2018assessing,dnmArchives} (updated regularly) where each good is recorded with its buyer, price, and time information. Identifying these payments on the Bitcoin blockchain is known as transaction fingerprinting~\cite{fleder2015bitcoin}. Darknet market vendors accrue big amounts of coins in multiple addresses~\cite{tai2019adversarial}. 

\subsection{Malware Payments}
 
Two types of malware are used to gain dark coins: {\it leakware} and {\it ransomware}.
\begin{itemize}
\item {\it Leakware} is an extortion tool, which hackers use to threaten a person with disclosure of private images or video recordings. 
\item {\it Ransomware} is a type of malware that infects a victim's data and resources and demands a ransom to release them. Once the ransomware is installed, it communicates with a command and control center. Although earlier ransomware used hard-coded IPs and domain names, newer variants may use anonymity networks, such as TOR, to reach a hidden command and control server. In the case of asymmetric encryption, the encryption key is delivered to the victim's machine. In some variants, the malware creates the encryption key on the victim's machine and delivers it to the command center. Once resources are locked or encrypted, the ransomware displays a message that asks a certain amount of cryptocurrency to be sent to an address. This amount may depend on the number and size of the encrypted resources. 
\end{itemize}

After payment, a decryption tool is delivered to the victim. However, in some cases, such as with WannaCry, the ransomware contained a bug that made it impossible to identify who paid a ransomware amount.  
 
\subsection{Undisclosed Transactions}

Dark coins may be received from undisclosed sales of goods or services. Typically, the coins are accrued to be spent on the blockchain without exchanging them for fiat currency. 

Traditionally, a notable source of undisclosed financial transactions has been the Hawala network which allows transferring money through personal connections globally~\cite{el2003informal}. Bitcoin can be used to facilitate transfers between trustless hawalas as follows. John gives \$100k USD to a hawala to transfer the equivalent amount to a location. A second hawala in the location receives bitcoins in its address and pays several dollars to John's friend Jim. Both hawala take a fee to allow the dollar transfer. John pays both bitcoin and hawala transaction fees but avoids verifying his identity or paying tax for the transferred amount. Hawalas undertake the risk of being discovered by financial authorities.

\section{Cryptocurrency Transaction Models} 

We follow the adversarial model in~\cite{androulaki2013evaluating} and observe the Bitcoin blockchain, denoted by txChain, for a period of time $\Delta_t$. In this period $\Delta_t$, n entities, $U = {a_1, a_2, \ldots, a_{nu}}$, send or receive transaction outputs in txChain through a set of $n_{\alpha}$ addresses: $A = {a_1, a_2,\ldots, a_{n\alpha}}$ where $n_{\alpha} \geq nu$. 
We assume that within $\Delta_t$, $n_\mathcal{T}$ transactions have taken place as follows: $T = {\mathcal{T}_1(I_1\rightarrow O_1),\ldots, \mathcal{T}_n(I_{n\mathcal{T}}\rightarrow O_{n\mathcal{T}})}$  denotes a transaction with sets of outputs and inputs, respectively. We will use $A(o_i)$ to denote the bitcoin amount in an output $o_i \in O$. Set of addresses in an output is denoted with $O.A$.

Cryptocurrencies have made a few design choices that facilitate money laundering. A fundamental problem stems from the Bitcoin transaction model, which has been adopted by its descendants (e.g., Litecoin) as well. A Bitcoin transaction lists one set of previous outputs  $I$, and one set of new outputs $O$, without explicitly stating which input is directed to which output (see Figure~\ref{fig:ransom}). Clustering heuristics assume that all inputs of a transaction belong to the same bitcoin user because they must be signed with private keys~\cite{meiklejohn2013fistful}. Outputs are not signed in the transaction, hence their owners are unknown.

With txChain, one can use the well-known multi-input, transition, and change address heuristics to link addresses to users~\cite{meiklejohn2013fistful}. However, the heuristics are error-prone. 

The Bitcoin transaction model has been adopted and extended in Monero and Zcash, which are called privacy coins. Monero modifies the input set by adding 10 decoy inputs for each true  input~\cite{moser2018empirical}. Zcash adopts Bitcoin's model for its \textit{unshielded transaction pool}, but also uses a \textit{shielded pool} where every transaction detail is cryptographically hidden~\cite{kappos2018empirical}.  

In our analysis we use the term {\it chainlet} to refer to bitcoin transaction substructures~\cite{akcora2017chainlet}. A first order chainlet is of type $C_{x\rightarrow y}$ which denotes a single transaction of $\left|I\right|=x$ and $\left|O\right|=y$. Higher order chainlets encode larger subgraphs.

We propose the following three criteria to quantify user, address, and transaction privacy on Bitcoin. 
\input{defin.tex}

\subsection{Datasets}

We download and parse the Bitcoin blockchain from 2009 to 2020. We use two types of data in our traceability analysis: darknet markets and ransomware payments. 
Our ransomware payment addresses are adopted from Montreal~\cite{paquet2018ransomware}, Princeton~\cite{huang2018tracking} and Padua~\cite{conti2018economic} studies which have analyzed networks of cryptocurrency ransomware. The combined dataset contains 24K addresses from 27 ransomware families. 

Our darknet market data is taken from DNM archives (\url{https://www.gwern.net/DNM-archives}). We use the grams dataset which contains web-crawled information about daily merchandise listings of 27 darknet markets between June 2014 and April 2016~\cite{dnmArchives}. Every listing has vendor, price in bitcoin, description, and country information.

\section{Bitcoin User and Transaction Privacy}

We explore two types of strategies to hide dark coin origins. Obfuscation strategies advise users to behave in certain patterns when dark coins are being spent. Evasion strategies recommend behavioral changes to avoid detection. 

\subsection{Obfuscation for Coin Movements}

Dark coins must be cashed out of blockchains by selling them on online exchanges for fiat currency. As online exchanges know the identities of customers and can report coin owners to law enforcement agencies, dark coins must first be passed through money laundering schemes where the taint is removed. 
Aware of identification risks, dark coin owners have used three money-laundering regimes since 2009 with increasing sophistication. 

In the first regime (since 2011), a high number of coins is passed through multiple transactions to hide origins (e.g. in peeling chains). That is, an observer is assumed to not have analytical tools to track the flow of coins in the large blockchain graph. With the increasing law enforcement activities and analytical capabilities, such obfuscation efforts prove futile.

Even when elaborate peeling chains are used, obfuscation efforts can be easily thwarted with analytical power and carefully selected features. For example, graph-based edge and address features can be extracted from the Bitcoin network to encode obfuscation patterns, cluster similar addresses, or even detect new dark coin payments~\cite{akcora2020bitcoinheist,weber2019anti}.

\vspace{2mm}
\noindent\textbf{Rule 1:} {\it Do not rely on transaction obfuscation to hide the origins of dark coins. Do not merge nor split coins by using elaborate chains. Avoid using an address to receive multiple payments.}

Since 2013, coin mixing~\cite{ruffing2014coinshuffle} schemes have been designed to further blur the flow, making coin tracking in the network a painstaking and often fruitless task. To involve ordinary addresses (i.e., untainted coins) into money laundering transactions, coin mixing uses coin payouts. Furthermore, mixing is repeated in multiple rounds with identical output amounts. At the end of the scheme, multiple addresses hold the laundered coins (minus the transaction and user payout fees), which can be sold on online exchanges separately. Once a high number of participants are found and multiple rounds are used, coin mixing provides very strong privacy. 
For example, in our analysis, we quickly reached 70\% of the daily Bitcoin network addresses by starting from known WannaCry ransomware addresses and including graph neighbors in two hops. As a result, online exchanges have started to outright deny buying or selling customer coins that are used in transactions that “appear like” mixing transactions. 

\begin{figure}
    \centering
    \includegraphics[width=0.95\linewidth]{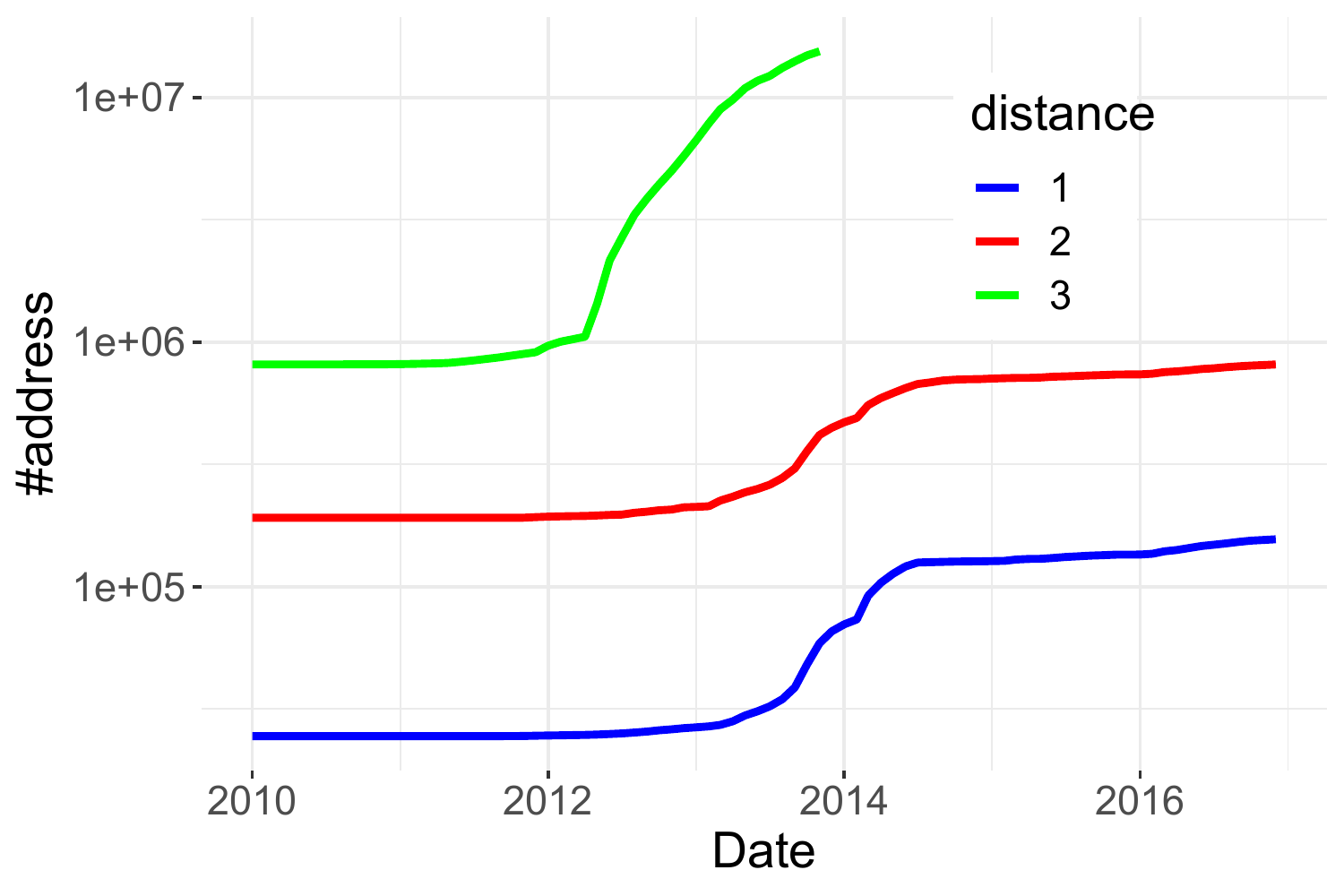}
    \caption{Graph distance from a known black (ransomware) address $a$ (i.e., number of addresses with $\mathcal{D}_a(k)=1,2,3$). For a distance of three or more, tracking suspicious addresses becomes untenable. }
    \label{fig:addresses}
\end{figure}

Figure~\ref{fig:addresses} shows the number of addresses that are reachable from known ransomware addresses. Ransomware hackers use coin-mixing transactions.  In 2011, we see the earliest ransomware addresses. 

For 1 and 2 distances, tracing cryptocurrency transfers are arduous because too many addresses are reachable from black addresses. For a distance of three, the number of suspicious addresses exceeds one million in a short time, which makes coin-mixing ideal for hiding dark coins.

\vspace{2mm}
\noindent\textbf{Rule 2:} {\it With multiple rounds in similar input-output amounts, coin-mixing allows enhanced security. However, exchanges will shun coins that exit coin-mixing rounds.} 

Shapeshifting (since 2017) is the latest money-laundering scheme on cryptocurrencies~\cite{yousafshapeshifting}. Tainted bitcoins are passed to an exchange where they are sold. The exchange pays the amount in coins of Monero, Zcash, or Dash cryptocurrencies. These privacy coins provide mechanisms to create transactions where input, output, and event amount information are hidden, which can be achieved in multiple ways. Dash uses a variant of coin mixing at the protocol level. Monero uses group signatures with ordinary past addresses as decoys in transaction inputs. ZCash uses zero-knowledge proofs and provides a shielded pool in which all transactions are hidden from the public. In privacy coins, transactions can be created to hide coin flow and eventually save coins in multiple addresses. The coins are again sold on the online exchange and the amount is paid as bitcoins. Coins (minus transaction and shapeshifting fees) are now back in the Bitcoin blockchain with no taint. To alleviate legal complications, a shapeshifting service (e.g., shapeshift.io) usually provides an API to track in-and out-flow of coins, which can be analyzed to link shapeshifting coins. Furthermore, users risk being identified when they use the shapeshifting website/service.

\vspace{2mm}
\noindent\textbf{Rule 3:} {\it Unless users make certain mistakes (e.g., returning to bitcoin immediately with very similar amounts~\cite{yousafshapeshifting}, shapeshifting can provide enhanced security for money laundering.}

\subsection{Evasion for Traceability Analysis}

Evasion must be considered at the protocol and blockchain levels. 

\subsubsection{Protocol Strategies}

The most basic evasion strategy is to avoid querying online API explorers, such as \url{www.blockchain.com/explorer}, with the address that holds your dark coins. It is safe to assume that search logs, along with your IP addresses, may leak to the public. 

\vspace{2mm}
\noindent\textbf{Rule 4:} {\it Do not query your address balance online.}

\begin{figure*}
\centering
\begin{subfigure}{.48\textwidth}
  \centering
  \includegraphics[width=.95\linewidth]{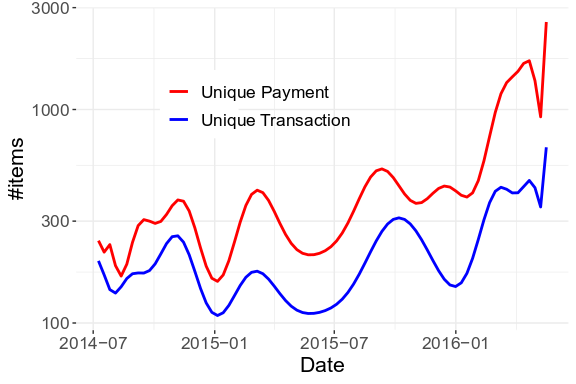}
  \caption{Number of items whose price match a single bitcoin transaction output of the day. Outputs can be a part of  $C_{N\rightarrow \leq 2}$ (Unique Payment) or $C_{N\rightarrow >2}$ (Unique Transaction) transactions. A single match may indicate that a merchandise is sold only once in a day.}
  \label{fig:vendorUnique}
\end{subfigure}
~
\begin{subfigure}{.48\textwidth}
  \centering
  \includegraphics[width=.95\linewidth]{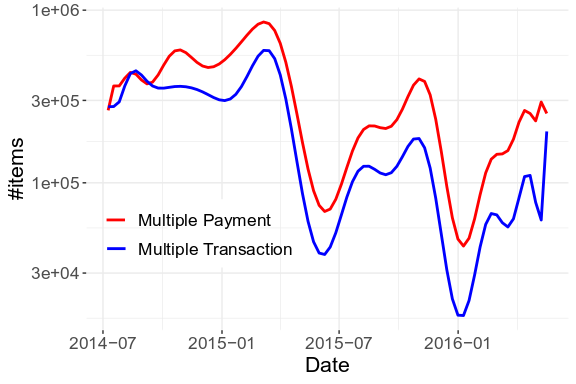}
  \caption{Number of items whose price match multiple  transaction outputs of the day. Outputs can be a part of  $C_{N\rightarrow \leq 2}$ (Multiple Payment) or $C_{N\rightarrow >2}$ (Multiple Transaction) transactions. Multiple matches may indicate that a merchandise is sold multiple times in a day.}
  \label{fig:vendorMultiple}
\end{subfigure}
\caption{Matching merchandise prices with bitcoin transaction outputs. We consider transactions of "payment" type as more likely candidates for darknet market sales.}
\label{fig:vendoramounts}
\end{figure*}

A second strategy is to hide your blockchain address from your wallet software, i.e., do not enter your private keys to your wallet. In a recent Monero and Zcash vulnerability, wallet software of a payee address could be triggered to send a message to the Peer-to-Peer blockchain network, which then revealed the wallet IP address to neighbors~\cite{tramer2020remote}. Such zero-day protocol attacks can never be fully eliminated. The solution is to parse the blockchain data with your code, which can be achieved  in 10 lines by using a Python library.\footnote{\url{https://github.com/alecalve/python-bitcoin-blockchain-parser}}

\vspace{2mm}
\noindent\textbf{Rule 5:} {\it Leave an air gap between your address and the web.}

When an output is spent, its associated address must be removed from the wallet. In \textit{dusting attacks}, an adversary sends a few satoshis to the address of an already spent output to locate new addresses owned by the same user. The attack benefits from a design mistake in Bitcoin wallets. Assume that an adversary knows that John used to own address $a_1$, and wants to learn his new address $a_2$ (if exists). Wallet software typically combines outputs from all user addresses to spend in a transaction. If John is not careful, the wallet will use the dust from address $a_1$ along with coins from $a_2$ as inputs to a transaction. By using the multi-input heuristic, the adversary will learn that John owns $a_2$. 

\vspace{2mm}
\noindent\textbf{Rule 6:} {\it Observe wallet behavior to detect obscure, unintended behavior. Similarly, do not accept default wallet behavior in transaction fee amounts. Wallet leaked data/metadata may facilitate linking your addresses.} 

A fourth strategy is related to hierarchical address creation, which generates a hierarchical tree-like structure of private/public keys and saves the user from having to generate multiple bitcoin addresses, one for each payment. The user has to store a \textit{seed} which is sufficient to recover all the derived keys and associated addresses. However, if the seed is leaked, an adversary can discover all user addresses and learn about their transactions on the blockchain. 

\vspace{2mm}
\noindent\textbf{Rule 7:} {\it Do not use hierarchically created addresses, which may be recovered even if you have deleted them from your wallet.}

\subsubsection{Blockchain Strategies}

Amount analysis on darknet markets shows that vendors make poor decisions in listing a price for their merchandise (we consider prices in bitcoins). Oddly specific prices with too many decimals, such as 0.067459 bitcoins, are very common in the dataset. This indicates that a vendor computes an exchange rate for a fiat currency amount and lists it as the merchandise price. However, Figure~\ref{fig:vendoramounts} shows that such an approach may aid transaction fingerprinting. In Figure~\ref{fig:vendorUnique}, we show that for every day we can match more than 300 merchandises to transaction payment outputs uniquely. Among these unique transaction outputs, most are part of a transaction that has one or two outputs (i.e., $\left|O\right|\leq 2$). Existing research approaches consider this pattern as spending or payment behavior\cite{androulaki2013evaluating}. In addition to the payment address, the remaining balance (if exists) is directed to a change or shadow address creating one or two outputs.

Figure~\ref{fig:vendorMultiple} shows merchandise fingerprinting results when a price matches to multiple transaction outputs. Multiple matching may be due to two reasons. First, the merchandise may have been sold multiple times. Second, the price may have matched non-sale related, ordinary bitcoin transactions. The second option can be optimized for an evasion strategy, which we discuss next.

\begin{figure}[t]
    \centering
    \includegraphics[width=\linewidth]{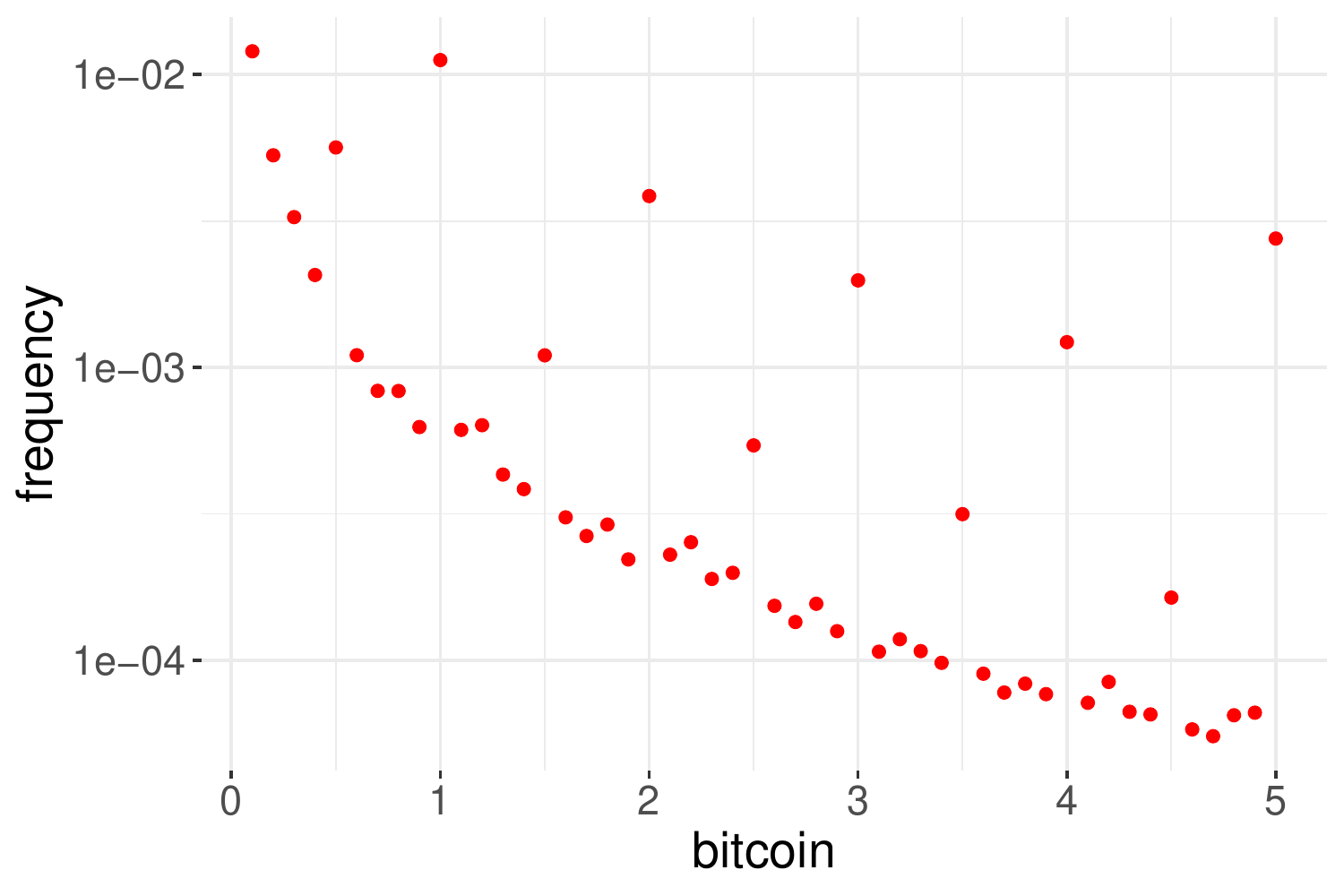}
    \caption{Bitcoin denominations ($i$) and their frequency ($N(i)$). When receiving a payment, amounts must be chosen with an eye to maximize matching transaction outputs.}
    \label{fig:amountfrequency}
\end{figure}

Figure~\ref{fig:amountfrequency} depicts bitcoin denominations and their frequencies in all bitcoin transactions. Multiples of bitcoin (e.g., 1, 2) are frequently found in bitcoin outputs. Popular amounts, such as 0.1 bitcoin, can be used as merchandise listing prices to fingerprint too many transaction outputs. With such frequent amounts, transaction fingerprinting of merchandise will be untenable for analysts.

\vspace{2mm}
\noindent\textbf{Rule 8:} {\it Avoid too specific bitcoin amounts, and use frequent denominations when receiving payments.} 

\begin{figure}
    \centering
    \includegraphics[width=\linewidth]{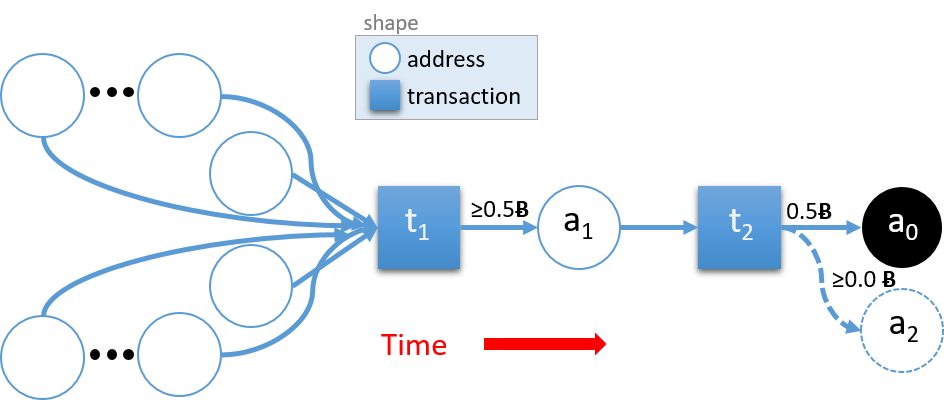}
    \caption{Ransomware payment to the black address $a_0$. Transaction $t_2$ usually contains a change payment as well (shown with the dashed edge to $a_2$.}
    \label{fig:ransom}
\end{figure}

The analysis shows that ransomed entities (e.g., companies, municipalities, hospitals) behave similarly when paying the ransom~\cite{akcora2020bitcoinheist}. First, the entity uses an online exchange to buy coins. The exchange facilitates this process by matching sellers to the coin buyer. An address $a_1$ is created for the entity and the bought coins are directed into it (Figure~\ref{fig:ransom}). As the ransom is usually a big amount (e.g., millions of dollars), the inputs of this transaction can be hundreds of addresses each contributing small coin amounts. This pattern is shown as the transaction $t_1$ in Figure~\ref{fig:ransom}, where the output amount is higher than the ransom amount so that a transaction fee can be paid in the ransom payment next. The transaction $t_2$ is the ransom payment. If there is a change amount left over from the ransom, it is directed to $a_2$. In 86.06\% of ransom payments, $t_2$ has one or two output address~\cite{akcora2020bitcoinheist}. An interesting fact is that the time difference between $t_1$ and $t_2$ is usually around 24 hours. This phenomenon implies that there exists a significant time gap between agreeing to pay and making a payment. 

 Once the coins reach $a_0$, hackers use money laundering methods to cash out. Hackers typically report $a_0$ to a victim to demand ransom but they do not control the payment pattern shown in Figure~\ref{fig:ransom}. Encoding this pattern in six features and searching the blockchain for the exact pattern matches, we find many true positives~\cite{akcora2020bitcoinheist}. The existence of such patterns, which the receivers choose to not control, facilitate payment detection.  
 
 \vspace{2mm}
 \noindent\textbf{Rule 9:} {\it Hinder traceability analysis by controlling the chainlet patterns used in preceding transactions.}

\begin{figure}
    \centering
    \includegraphics[width=0.95\linewidth]{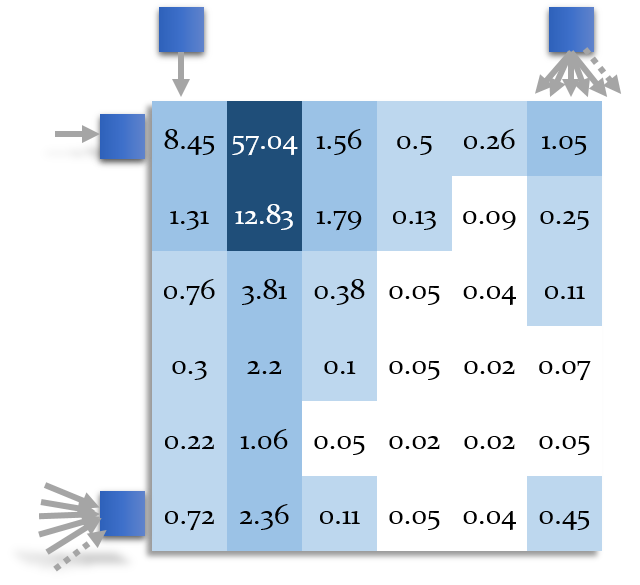}
    \caption{Percentages of chainlets (transactions with number of inputs (in rows) and outputs (in columns)) in a chainlet occurrence matrix of $6\times 6$~\cite{akcora2017chainlet}. Dark colors indicate higher values. 57.04\% of all bitcoin transactions have one input and two outputs (i.e., $\mathcal{C}(1,2)=0.5704$ of all transactions).}
    \label{fig:perc}
\end{figure}

Controlling for earlier chainlets can also be applied to darknet market payments. Figure~\ref{fig:perc} shows the percentages of chainlet types in the Bitcoin blockchain. Most transactions have two outputs (second column). When receiving a payment, the transaction can be created as a $C_{1\rightarrow 2}$ chainlet to maximize the number of matching transaction outputs.

\vspace{2mm}
\noindent\textbf{Rule 10:} {\it Payment transactions must consider chainlet frequencies to minimize traceability which is achieved by receiving payments through transactions of $\leq 2$ inputs and $\leq 2$ outputs.}

\section{Path to Deployment}
\input{path.tex}
\section{Conclusion}
Interdisciplinary methods at the interface of statistical data analysis and protocol software analysis play an increasingly important role in the detection of money laundering schemes and constitute an emerging research direction in Data Science. Although there have been attempts to identify users and transactions, we believe that the success of such works is due to user and protocol mistakes that can be easily avoided. 

In this paper, we have presented empirical results to argue that when a few rules are followed, address and transaction traceability efforts on blockchains become computationally prohibitive with an error rate too high to be feasible.

\fontsize{9.1pt}{10.1pt} 
\selectfont
\bibliography{thebes}
\end{document}

%% file: defin.tex
\vspace{2mm}
\noindent\textbf{Distance Anonymity. } For two addresses $a_k$, and $a_m$ and a transaction path
$T=\{\mathcal{T}_0(I_{0}\rightarrow O_{0}),\ldots,\mathcal{T}_n(I_{n}\rightarrow O_{n})\}$ 
where $\forall \mathcal{T}_i \in T|i> 0$, $I_i.A\cap O_{i-1}.A\neq \emptyset$,  $ a_m \in \mathcal{O}_0.A$ and  $a_k \in \mathcal{O}_n.A$, the distance of $a_k$ to $a_m$ is defined as $\mathcal{D}_{m}(k)=\arg\min_T({n})$. 

\vspace{2mm}
\noindent\textbf{Amount Anonymity:} For an output amount $A(o_i)$ and a time period $\Delta_t$, we observe $T=\{\mathcal{T}_0(I_{0}\rightarrow O_{0}),\ldots,\mathcal{T}_n(I_{n}\rightarrow O_{n})\}$ transactions where $\forall \mathcal{T} \in T$, $\exists o_j \in \mathcal{T}.O|A(o_j)=A(o_i)$. Amount anonymity $\mathcal{N}(i) = \left |T\right|$. 

\vspace{2mm}
\noindent\textbf{First-Order Chainlet Anonymity:}.  For a transaction with $i$ inputs and $o$ outputs and a time period $\Delta_t$, we observe $T=\{\mathcal{T}_0(I_{0}\rightarrow O_{0}),\ldots,\mathcal{T}_n(I_{n}\rightarrow O_{n})\}$ transactions where $\forall \mathcal{T}_i \in T$, $\left|I_i\right|=i$ and $\left|O_i\right|=o$. Chainlet anonymity $\mathcal{C}(i,o) = \left |T\right|$. Note that higher order ($k\geq 1$) chainlets can be used to control for larger subgraphs on the bitcoin graph.



%% file: path.tex
Our results suggest 10 rules to attack and render AI based cryptocurrency tracing tools ineffective. Although discouraging, the adversarial setting is a promising research domain for novel AI based models that can counter our evasion strategies.

In our analytics deployment, the main performance bottleneck has been the high level of connectivity in the Bitcoin transaction network, which makes graph computations untenable. Although bitcoin recommends six block confirmations for a transaction, we encounter a  set of rapidly moving bitcoins in the network. Many coins are mined (transacted) in every block of a 24 hour period. To avoid the connectivity problem, we use a time windowed approach and extract features from a limited depth network. 

A second issue is related to within-exchange transaction activity, which remains hidden from the blockchain. In the same vein, second layer solutions, such as Lightning Network, leave most transactions off the main blockchain, and complicate traceability issues greatly. AI deployment efforts must develop and deploy models that combine on and off-the-chain transactions and produce probabilistic traceability models. 